\documentstyle[emulateapj,epsfig,apjfonts]{article}
\def\gs{\mathrel{\raise0.35ex\hbox{$\scriptstyle >$}\kern-0.6em 
\lower0.40ex\hbox{{$\scriptstyle \sim$}}}}
\def\ls{\mathrel{\raise0.35ex\hbox{$\scriptstyle <$}\kern-0.6em 
\lower0.40ex\hbox{{$\scriptstyle \sim$}}}}
\lefthead{Chapman et al.}
\righthead{Radio galaxy selected galaxy clusters at high redshift}

\begin{document}
%\doublespace

\title{Radio galaxy selected clusters at high redshift 
and associated ERO overdensities}

\author{S.\,C.\ Chapman,$\!$\altaffilmark{} P.\ McCarthy,$\!$\altaffilmark{}
\& S.\,E.\ Persson\altaffilmark{}
}
\affil{Observatories of the Carnegie Institution of Washington,
Pasadena, CA 91101,~~U.S.A.}

%\setcounter{footnote}{5}

%\altaffiltext{2}{something}

\begin{abstract}
Galaxy clusters at high redshift present a superb opportunity to study
galaxy evolution, with large galaxy samples at fixed distances.
Several lines of evidence point towards
large formation redshifts and passive evolution for cluster ellipticals.
At redshifts larger than $z\sim0.8$,
this picture rests on data from only a handful of rich clusters.
We have observed four potential clusters
%the hosts of radio galaxies at known redshift, 
to sample sparser environments in a
critical redshift range, $0.8 < z < 1.2$.
%and will complement studies of
%X-ray and optically selected clusters.
%In particular, constraining cluster membership through photometric
%redshifts allows 
We compare the photometric
evolution and radial density profiles
in our clusters with more massive clusters
studied in a similar manner.
%A radial density analysis is also used to 
%to measure the blue galaxy fraction.
We also highlight the overdense ERO 
($I-K>5$) population, significantly redder than
the old elliptical population at the cluster redshift.
We discuss the implications of our clusters at high redshift,
and possible differences in environment from richer clusters at similar
redshifts selected through X-ray or optical techniques.

\end{abstract}


%\newpage

%
%
%
\section{Introduction}

High redshift galaxy clusters present luminous signposts for the formation and
evolution of massive galaxies. While the study of field galaxies at
very large redshifts ($z > 3$) has recently progressed ({\it e.g.}
Steidel et al.~1999), clusters continue to
be our best probe of dense environments and large mass scales at
early epochs.
Clusters with luminous AGNs at their centers offer a different
perspective on galaxy interactions and activity than clusters
selected either on the basis of optical over-densities or spatially
extended  X-ray emission.
The AGN-selected clusters
tend to be poorer than the optically and X-ray selected rich clusters that
dominate most catalogs, and thus probe a different range of environments
(Lilly \& Prestage 1987).
Such a sparse environment may also be related to the formation and fueling
of the luminous AGN (Heckman et al.~1986).
%since rich clusters tend not to have AGN dominated
%brightest cluster galaxies (BCMs). ?????????

The existence of massive collapsed objects, particularly rich galaxy clusters,
at high redshifts provides a challenge for theories of cosmic structure
formation (e.g. Press \& Schechter 1974). The z$\sim$1 range represents a
critical point at which serious constraints can be placed on the formation
and evolutionary history of early-type galaxies in clusters.
Beyond z$\sim$1 in cosmologically flat CDM models, the amount of
merging occurring within the prior $\sim$1 Gyr was thought to be
 sufficiently large so
as to dramatically inflate the locus of early-type galaxy colors.
Recent modelling has shown however that the locus of such red galaxies
may remain quite tight until beyond $z\sim1.5$ (Kauffmann \& Charlot 1998). 
However these models still have difficulty predicting the correct slope of
the red sequence.

The emerging evolutionary picture from the study of a few high redshift galaxy
clusters (e.g. Ellis et al.~1997,
Stanford et al.~1998, Kodama et al.~1998)
is one of smooth and steady spectro-photometric evolution for early-type galaxies over more
than half the age of the universe - implying large formation redshifts and primarily passive evolution.
The average E+S0 galaxy color at a particular luminosity evolves
gradually, while both the slope and scatter of galaxy colors around the
mean change very little.
The small scatter implies that, even approaching z$\sim$1, we have not
reached the time at which a significant fraction 
%*the bulk* is a misnomer since it takes relatively star pop to affect color
of the stars in early-type galaxies were young.
The fact that the color-magnitude slope does not change with
redshift
strongly favors a scenario in which this slope arises from a mass-metallicity
correlation (Stanford et al.~1998). 
%since more massive galaxies would retain SNe ejecta
%more effectively from the initial burst. 
Different burst ages and epochs
would result in a changing slope with redshift.

The evolution of luminous AGN may be tied to the evolution of their parent 
environments, particularly in clusters. The peak in the QSO luminosity 
evolution is thought to have occurred
between 1$<$z$<$3 (e.g. Boyle \& Terlevich~1998).
This was {\it the age} of quasars.
In the present epoch, powerful AGN are
far less numerous and appear not to inhabit very overdense regions. However,
prior to z$\sim$0.5 conditions were different,
many radio-loud AGN existed in overdense or cluster environments
(e.g. Hall et al.~1998, Yates et al.~1989, Yee and Green 1984; Hill \& Lilly 1991; Ellingson, Yee, \& Green 1991).
If mergers and interactions are indeed the key to fueling
luminous AGN, then the temperatures and velocity dispersions in later-epoch
clusters ($z\sim0$) may be too large for efficient mass transport. The
strong QSO evolution may be naturally explained as a result of cluster
virialization and evolution -- at higher redshift, the lower galaxy
velocities lead to higher interaction probabilities (e.g. Fabian \& Crawford 
1990).
A luminous AGN present at the cluster center may be an indicator of rapid
dynamical evolution of the cluster and the constituent galaxies.

Although a few recent studies have addressed cluster elliptical evolution at
redshifts $\sim$1
(Stanford et al.~1998, Stanford et al.~1997, Rosati et al.~1999,
Tanaka et al.~2000, Kajisawa et al.~2000),
there is a marked paucity of data available for clusters
in this critical redshift regime (z$\sim$1), especially those with less
rich environments where mergers are expected to be more efficient.
We have identified a sample of potential high redshift clusters from
a complete sample of radio galaxies (Chapman et al.,
in preparation), the  
Molonglo Reference Catalog 1 Jy sample (McCarthy et al. 1996; Kapahi et al. 1998).
We have selected all sources identified with galaxies with measured redshifts in the 
range $0.8 < z < 1.5$, using visible and
$K$s-band images obtained with the
2.5\,m du~Pont telescope at Las Campanas.
Since old elliptical cluster galaxies have very red colors, $K$-band
galaxy selection removes a large source of blue interloping field
galaxies, highlighting potential cluster members. 
In addition, cluster members do not suffer as greatly
from k-corrections in the $K$-band as in the optical (Stanford et al.~1998)
and $K$-magnitude/color diagrams therefore
present a comparable study of cluster evolution across a range of
redshifts. 

In this paper we present deep multiband images of four obviously
overdense  radio galaxy (RG) fields from our sample at
$z>0.8$.
Confirming the cluster status will
ultimately require spectroscopic measurements of redshift for many cluster
members. With the present data, we rely on Color-Magnitude diagram (CMD) 
analyses
to reveal the characteristic red sequence of elliptical cluster members,
in consort with the measured radio galaxy redshift, and in one case (MRC\,0959-263)
a few cluster member redshifts lying fortuitously along the slit (sections 3.1,3.2).
A radial density analysis is also used to assess the cluster richness over
the field population (section 3.3). 
%as well as to measure the blue galaxy
%fraction (section 3.3).
We also highlight the overdense ERO population, significantly redder than
the old elliptical population at the cluster redshift (section 3.4).
In section 4, we discuss the implications of our clusters at high redshift,
And possible differences in environment from richer clusters at similar 
redshifts selected though X-ray or optical techniques.

\section{The Sample, Observations and Reduction}

The four radio galaxy fields in this study were chosen from a large sample
of $K$- and $V$-band images, because of an apparent overdensity
of red galaxies in the $K$ image when compared with $V$. 
We also required that the ($V-K$) vs $K$ CMD revealed a linear sequence of red
galaxies including the radio galaxy as further evidence for the cluster
environment.
This was not meant
to be systematic for the sample, but merely to choose a few potential clusters
at $z\sim1$ for deeper followup observations.

\subsection{Imaging}
The optical and near-IR images on which our present analysis is based 
were obtained using
the du~Pont 100\,inch telescope at Las Campanas, Chile, using 
CCD cameras in the optical. The $J$ and $H$-band images were obtained with
the CIRSI camera (Beckett et al.~1998 -- 1k$\times$1k pixel Hawaii array), while the $K$s-band
images were taken with IRCAM (Persson et al.~1992 -- NICMOS3 256$^2$ array).
The details of the optical
observations and their reduction can be found in 
McCarthy et al.~(1996). The near-IR observations were taken with a 
30\arcsec\ dither pattern, and reduced in a standard approach for IR
imaging.

To facilitate the analysis of the datasets, the various images of the
four cluster 
fields were resampled and rotated to common reference frames using
the IRAF tasks GEOMAP and GEOTRAN. The fields are centered on the initial
estimate of the cluster positions from the red galaxy
spatial distribution. 
The final images have the same pixel size as the 0.265 arcsec/pixel 
CCD images.

The source extraction software SExtractor (Bertin \& Arnouts 1996) was
used to make an initial object detection pass at $>$3$\sigma$
with the final catalog at $>$4$\sigma$
in the $K$s-band images, while 
measuring flux parameters for each individual passband in the separate
images. Magnitudes and colors were measured using a 2 arcsec diameter
aperture, roughly twice the FWHM of the images.
The depths reached in each filter vary somewhat field to field, but are
typically close to 4$\sigma$ detection limits of $K$s=20.5, $H$=21.4, $J$=21.4,
$I$=24.0, $V$=25.0. The actual limits for each field are presented in
Table~1. 
Figs.~1 and 2 present the $V$ and $K$-band images of the four clusters,
while Fig.~3 depicts a false color image of the richest cluster in our
sample, MRC~1022-299.
For visualization, an $I-K$ color-coded representation of each 
field is presented in Fig.~4, with red squares highlighting probable
cluster members (those lying within the presumed locus of early-type galaxies
as discussed in the next section).

The Galactic $E(B-V)$ toward the four fields is small ($\approx 0.04$), 
and we have not applied an extinction correction to our photometry.
We estimate the quadrature sum of the
systematic 1$\sigma$ errors in the photometry (zeropoints and PSF matching) 
to be $\sim$0.05 for the IR colors, and
$\sim$0.04 for the optical-$K$ colors.

\subsection{CTIO 4m spectroscopy}

The RC spectrograph on the CTIO 4m telescope
was used to obtain optical spectra for the four radio galaxies,
although no clear redshift was measured for MRC~0527-255.
For MRC~0959-263, two other galaxies falling on the slit had redshifts
within 5\% of the radio galaxy, and are likely cluster members.
Details of the observations and reductions can be found in
McCarthy et al.~(1992), while the
spectra will be presented in McCarthy et al.~in preparation.

%Compute the cluster center from the brightest members of red sequence

\section{Analysis and Results}

\subsection{Color-magnitude diagram analysis}

The color-magnitude diagram (CMD) forms our main diagnostic for identifying
red galaxy overdensities, and we present a brief overview of the 
technique before discussing each cluster individually.
The CMD for a galaxy cluster typically reveals a
tight red sequence of massive elliptical galaxies, implying a mass-metallicity
relation for early-type cluster galaxies (Stanford et al.~1998, Gladders et
al.~1998).
This argument can be inverted to
search for the presense of a cluster via the red elliptical sequence.
The fact that studies to date have implied large formation redshifts
and primarily passive evolution
for the cluster ellipticals allows the red sequence to be used as
an estimate of the cluster redshift, even at redshifts out to $z\sim1$.
We have used this approach to characterize the environments of the four
radio galaxies under investigation here.
In the cases where the radio galaxy redshift is known, we have, along with
$z_{CMD}$,
two independent measures of the cluster redshift to compare.

Figure 5 %~\ref{fcmd} 
shows the CMDs for the RG fields
in observed-frame colors. % $V-K$, $I-K$, $H-K$
For the two higher redshift clusters, $J$- and $H$-band images were obtained
since the near-IR %$J-K$
colors begin to show a significant deviation from the field population.
Two radial cuts have been made on the object catalog centered on the
radio galaxy,
with solid squares representing objects
within 35 arcsec %(1/3 Abell radius at $z\sim1$)
of the radio galaxy, crosses from 35\arcsec\
to 60\arcsec\,
and circles for objects greater than 60\arcsec\,.
The solid diagonal lines represent $4\sigma$ detection
thresholds in color.
A color-magnitude sequence of red galaxies, which includes the
radio galaxy, is visible in
Fig.~5 for all clusters, with many of the red objects lying within the
central 35 arcsec. 
Our position-color representations of the cluster fields (Fig.~4)
depict the red sequence objects with red squares, while 
blue triangles depict bluer objects
lying below the red sequence. 

%The sequence is tighter for the redder colors%except 0527?
%more apparent in richer cluster 1022

Plotted in all
panels of Figure 5 %\ref{fcmd} 
(dashed lines) are estimates of the no--evolution color--magnitude locus
for early--type galaxies.  
The dashed lines are derived from
the dataset described in Stanford et al.~(1998) which
uses multiband %$UBVRIzJHK$ 
photometry of early--type galaxies
covering the central $\sim$1 Mpc of the Coma cluster (Eisenhardt et al.~1997, 
Stanford et al.~1998).  The dashed lines therefore represent the colors
that Coma galaxies would appear to have if the cluster could be placed at
the redshifts of our radio galaxies
and observed through the $VRIJHK$s filters we used.
The dashed lines were interpolated from the values presented in
Stanford et al.~(1998), where they have a cluster very near the redshift
of each cluster in our sample to $z=1$.
For MRC0527-255, we made a very small extrapolation to the no-evolution Coma
line from the Lynx field clusters at $z=1.26$ presented in Rosati et al.~(1998),
and Stanford et al.~(1997).

Our procedure was to fit a line to all objects within $18<K<20$, taking
   a color cut of $I-K\pm0.4$ around the Coma no-evolution line 
   derived for the
   radio galaxy redshift. For this line we then determined the equivalent
   Coma no-evolution redshift. 
   In fitting the red sequence in this manner, we noticed that
    measuring the slope and scatter of these objects
   is plagued with systematic errors, since the values changed considerably
   depending on the $I-K$ region used to fit over, with no objective criterion
   to select the $I-K$ fitting region.
For the one RG without a redshift, MRC\,0527-255, 
the dashed line represents a fit
of the no-evolution Coma colors to the
brightest red galaxies including the RG, within
the central 35 arcsec, satisfying the color cut of $3.6 < I-K < 4.4$.
%additional fit? $2.7 < I-K < 3.3$.
The results are presented in Table~1, and compared to the
measured redshifts of the radio galaxy in the cluster.
The uncertainty in the derived redshift results from
neglecting to correct for Galactic extinction, from the systematic
zero-point uncertainties in the Coma data (which we take as 0.075 mag
-- Stanford et al.~1998), the photometry, and the choice of objects
used for the fit, giving rise to a $\sim0.10-0.15$ mag error for the
individual clusters.
The $I-K$ red sequence shows a much tighter relation than $V-K$, likely
because the extinction is less and the effects of any recent star formation
are less pronounced at the longer wavelength.

The case for MRC~0527-255 is less clear than the other three.
The $I-K$ red sequence containing
the radio galaxy indicates a redshift of $z\sim1.3$, which would explain
the difficulty in measuring a redshift for the radio galaxy with CTIO~4m
blue spectrum.
Spectra of the brighter cluster members will be required to verify whether
this indeed represents such a high-$z$ cluster.

Certainly for the three radio galaxies with measured redshifts, the location
of the expected locus of Coma colors is remarkably consistent with the
colors of the red excess galaxies in the field. 
%Even with interlopers
%possibly increasing the scatter of objects around this sequence, we find
%no obvious evidence for color or slope differences from the
%no--evolution Coma colors, although
MRC~0959-263 and MRC~1139-285 do show small color offsets from
those expected from the RG redshift (see section 4). 
The color difference is measured by the difference at $K=19$ between the Coma
no-evolution line at the RG redshift and our red sequence fit line.

%or is there a problem, with some? Lack of redshift identification makes
%it difficult to estimate

%lack of morphology selection makes color scatter comparisons difficult

\subsection{Individual Clusters}

In this section, we present details of the individual radio galaxy (RG) fields
in order of increasing redshift. 
%Properties of the cluster fields are summarized it Table~1.
%radio properties, luminosities, cluster compactness.

\subsubsection{MRC\,0959-263}

%This field was observed in $K$,$I$, and gunn-$r$, to 4$\sigma$ depths of
%20.5, 24, and 25 respectively.
The $K$=17.7 radio galaxy lies near the bright end of the red sequence of 
galaxies, but a $K$=17.0 elliptical lying 20.7\arcsec\ to the northwest
may actually be the {\it brightest cluster member} (BCM) galaxy.
The RG has a redshift of z=0.68, but the extrapolated redshift
from the red sequence is higher at z=0.75. %This is significant since
The no-evolution $I-K$ color difference is $\sim$0.15$\pm$0.10 magnitudes.

\subsubsection{MRC\,1139-285}

%The 4$\sigma$ depths reached for MRC\,1139-285 were $K=20.7$, $I=24$, $V=25.5$.
This radio galaxy appears to be a likely candidate for the BCM, 
situated as the brightest galaxy on the red sequence. Another bright red
galaxy lying next to the RG on the CMD diagram is actually a barely
resolved pair lying 15\arcsec\ to the northwest.
Although the RG has a redshift of z=0.85, the cluster sequence shows
very similar colors to MRC\,0959-263, with an estimated redshift of
z=0.77. This is not very significant against the no-evolution $I-K$
color difference of $\sim$0.18$\pm$0.11 magnitudes.
The cluster members are much more spatially dispersed than in 
the other three fields, with less
obvious central overdensity surrounding the RG.

\subsubsection{MRC\,1022-299}

A false color image of MRC\,1022-299 (Fig.~3, plate 1) shows the 
dramatic overdensity of red galaxies, the most obvious overdensity from our
sample of four radio galaxy fields.
This is also apparent in the clear pileup of galaxies along the red 
sequence in the color magnitude diagram (Fig.~5).
%The 4$\sigma$ depths reached were $K=21.0$, $H=21.7$, 
%$J=22.4$, $I=24$, $V=25.5$.
The red sequence, with an estimated
redshift $z=0.95$, clearly includes the RG. The cluster redshift, 
estimated from the red sequence, is slightly
higher than the spectroscopic redshift of the RG itself ($z=0.93$). 
The radio galaxy is not the BCM, with the actual BCM on the red
sequence ($K$s=17.8) lying 38\arcsec\ from the RG.
%[QUES: MORE COMMENTS ON THIS LARGE SEPARATION FROM THE BCM?]

\subsubsection{MRC\,0527-255}
%This field represents something of an enigma.
Although no redshift was measured for for the radio galaxy MRC\,0527-255,
the $K$-redshift relation for radio galaxies (e.g. Lilly~1989,
McCarthy~1993) and the $8{''}$ apeture K magnitude of 17.2 implies
$z\sim1.2$, a
higher redshift than the other radio galaxies in the present sample. 
However, the $K-z$ relation
has significant scatter, (dz$\sim$0.3 RMS at $z=1.2$), 
and it would not be
surprising to find the RG and its associated over dense environment at a
somewhat lower redshift.

The radio galaxy is amongst the brightest members of a loose
sequence of galaxies with very red colors, where the estimated sequence
redshift would be $z=1.28$. Comparison with the two clusters at similar redshift
discovered by Rosati et al.~(1999) and Stanford et al.~(1997) show very
similar colors in all bands, lending plausibility to this field representing
a high redshift ($z>1$) cluster.
%There is also some evidence for a red sequence of galaxies at $z\sim0.75$,
%and there may be a secondary cluster
%structure projected along the line of sight.

In Fig.~6 % \ref{0527photz}, 
we plot the measured flux converted to $AB$-magnitudes
in all known bands of the
six $K$-brightest objects lying on the red sequence,
within the central 35\arcsec\ radius of the RG.
The $K$-band flux for all galaxies has been normalized to that of the
RG. Overplotted are 
Coleman, Wu \& Weedman (1980) %Bruzual \& Charlot (1997)
elliptical galaxy models redshifted to $z = 0.8$, $1.3$ and $2.0$.
%with passive evolution of a 1 Gyr burst stellar population with solar 
%metallicity formed at $z = 5$ for $h=0.65$, q$_0 = 0.1$.
The templates are at least suggestive that if these objects are old cluster
members, they lie at redshifts greater than $z=1$, and are
consistent with a cluster lying at $z=1.3$.

\subsection{Radial Density}

Having identified apparent red sequences from the CMD 
in all four candidate clusters, 
we can explore the density of the fields as a function of radius from the
radio galaxy. 
The galaxy density was calculated in successive annuli 5\arcsec\ wide for each
cluster and for the average of all four clusters together 
(Fig. 7a). %~\ref{density}a). 
Figure 7a %~\ref{density}a 
clearly shows a significant excess above the field galaxy
limit at our survey depth, $K$s=20.5 
(solid horizontal line with 1$\sigma$ errors due to variation in 
counts across the field as dashed lines - 
Cowie et al.~1994,1996 and Saracco et al.~1999) for the 
four clusters averaged together. 
%Error bars are
%calculated for the number of objects in each bin as per Gehrels (1986).
The average integrated counts over the central 25\arcsec\ radius are
$>6\sigma$ above the field population, estimated conservatively as the 
field count plus 1$\sigma$.
However the signal in the core region
for any cluster on its own is still convincing. 
% EBARS USED TO BE $\pm 1\sigma$ of the individual cluster surface densities.

By taking an $I-K$  color cut on the objects of 1 magnitude
bluer than the red 
sequence in each case (Fig.~7b), %\ref{density}b 
the significance of the overdensity
signal becomes larger, as we have effectively removed a portion of the
contaminating field galaxy population. 
(peaking at 65\,arcsec$^{-2}$ $\approx$ 6$\sigma$
above the field)
However, such a cut also removes
all the blue galaxy population residing in the cluster which was 
bright enough to be detected at $K$s-band.
Cowie et al.~(1996) present the counts of field objects
of the Hawaii deep surveys with various
 $I-K$ color cuts. In 
Figure~7b %\ref{density}b 
the field levels are estimated for each cluster using the $K$s=20.5 count
with an  $I-K$ color cut one magnitude bluer than the red sequence value
at $K$s=19 -- the central value of our fitting procedure described in 
section 3.1.
The $I-K$ field limits vary for the clusters 
from $\sim6$/arcmin$^2$ (MRC\,0959-263) to $\sim3$/arcmin$^2$ (MRC\,0527-255).
The average integrated counts of the respective $I-K$ 
field limits over the central 25\arcsec\ radius is a 7.5$\sigma$ detected
overdensity.

There appear to be two spatial components to the overdensities in all
clusters (Fig.~7).
The first is an overdensity within $\sim$30\arcsec\ of the RGs 
relative to our
own data at $>$30\arcsec\, (a {\em near-field} excess). 
For all objects with $K$s$<20.5$ this
represents a 4.5$\sigma$ integrated excess
averaged over the four clusters. 
However, even excluding this {\em near-field} excess, the integrated counts
for all clusters at radii $>$30\arcsec\ relative to
random-field counts from the literature (Cowie et al.~1996,
Saracco et al.~1999) represent a 4.5$\sigma$ detection.
This {\em far-field} excess 
appears to extend to $>$80\arcsec\,, although our data are sparse at these
radial distances.
%The near-field excess is present in $\gtrsim$25\% of the fields, and the
%far-field excess in $\gtrsim$50\%.
The amplitude of the near-field excess is insufficient to explain the total
excess $K$s-band counts, and the far-field excess is likely real.
The overdensity in our four fields corresponds to an approximate Abell
class of 1.5$\pm$0.5, where the red sequence extrapolated redshift is used
in all cases to measure the angular scaling.

%The nearest cluster, MRC~0959-263, shows a more general overdensity out to 
%80\arcsec\ radius, possibly related the 
%consistent with it's expected larger size, and 
%greater effective depth reached at the cluster redshift ($z\sim0.68$).
%MRC~0527-255, shows a 
%compact core ($<20$\arcsec\,), 
%consistent with having a large redshift ($z\sim$1.3).

\subsection{EROs}

Radio galaxy and quasar fields have shown apparent overabundances of EROs
(e.g. Liu et al.~2000, Dey, Spinrad \& Dickinson 1995, McCarthy et al.~1992). 
The most natural explanation is that these might represent
evolved early type galaxies in a cluster or group environment.
Without morphological or sub-mm information, however, it is difficult to know 
if any might represent dusty obscured starforming objects 
(Dey et al.~1999, Cimatti et al.~1999).
Although all those objects with $I-K > 4$ would be considered anomanously red
if discovered in the field, in high-$z$ cluster fields they lie close
to or along
the red galaxy sequence, and their apparent overdensity in our
four clusters is large, especially in the two higher-$z$ fields.
However, the objects with $I-K\gtrsim5$ are clearly too red to form part of the
typical mass/metallicity related sequence of cluster members. 

We define two samples of EROs in our RG fields: those with $I-K>5$,
and those objects which are significantly above the red sequence
($I-K>4.3$ for the lower-$z$ clusters, $I-K>4.7$ for the higher-$z$ clusters).
Although many such objects are detected,
very few of these objects have $K<19$ and a direct comparison 
with the random field ERO($R-K>6$) population of Thompson et al.~(1999)
is difficult.
In Table~1, we list the ERO($I-K>5$) and ERO($I-K>$red seq.~) 
numbers for each cluster to limits of $K<20.0$ (S/N$>5\sigma$) and 
$K<20.7$ (S/N$>3.5\sigma$). 
As some of the  candidate
EROs appear near the edges of the field, where the exposure
time is less than in the central regions, they do not meet our
detection criterion of $5\sigma$ and $3.5\sigma$ for the two respective
samples, and we do not include these in our catalog. 
For ERO($I-K>5$), we find %12 and 25 
10 and 18 objects in the $K<20.0$ and $K<20.7$ samples respectively,
while for ERO($I-K>$red seq.~), we find %16 and 35
14 ($K<20.0$) and 30 ($K<20.7$)
objects. The number of EROs is similar in the four clusters.
As the optical magnitudes are very faint in many cases
(below our nominal 4$\sigma$ detection threshold), all the ERO 
candidates with $K$s$<20$ were verified to be present 
by visual inspection of the multi-filter data.
Several of the $K<20.7$ sample were only detected at $K$s-band.

A field survey 
using HST-NICMOS data (Yan et al.~2000) derives surface
densities of 0.31$\pm$0.14 arcmin$^{-2}$ and 0.63$\pm$0.2 arcmin$^{-2}$, 
with $R-H\gtrsim5$ and $H<20.0$, $H<20.5$ respectively. Yan et al.~(2000)
liken their $R-H>5$ to $R-K>6$, which has been shown to be roughly equivalent
to our $I-K>5$ ERO criterion (Liu et al.~2000).
Comparing directly our ERO($I-K>5$) samples 
to Yan et al.~(2000), we find surface densities of
$0.78\pm0.25$ arcmin$^{-2}$ (2.5$\times$ the field)
and $1.41\pm0.34$ arcmin$^{-2}$ (2.2$\times$ the field).

%%%%% ADD LATER PERHAPS %%%%%%%%%
%We note that the {\it field} density of EROs is a difficult number to 
%quantify, since they appear to cluster quite strongly (McCarthy et al.~2000). 
%The Yan et al.~(2000) sample drops in density by 25\% to 40\% when apparent
%ERO clusters are excluded. Nonetheless, our 4 cluster fields represent
%a fairly small area survey, and only extreme serendipity could result
%in such an apparent ERO overdensity in each of the cluster fields.
%%%%%%%%%%%%%%%%%%%%%%%%%%%%%%%%%%

The EROs are not generally within the $\sim35$\arcsec\             
near-field overdensity region, and tend towards the outer regions of
our observed fields (with the possible exception of MRC~0527-255).
Our candidate EROs are represented as black circles in the
position-color plots (Fig.~4).
This may be indicative of the formation processes that 
lead to the extremely red colors if the EROs are
related to the high-$z$ cluster environment.
%field EROs likely signify a cluster mass sometimes
%Or do they mostly appear overdense in RG environments because of
%membership in the cluster itself as passively evolved old ellipticals.
A small percentage of the already overdense
old and red cluster galaxies might have enough dust to further 
redden their colors
significantly above the mass/metallicity sequence. 
Alternatively, there may be an overdensity of dusty 
ULIRG-type galaxies formed
in the %dense, compact 
dense environment of the radio galaxy.
Their location in the
outer field excess would be consistent 
with objects undergoing some phase of merger (e.g. van~Dokkum et al.~1999).
%objects which may not yet be tidally stripped of their gas and dust
Sensitive sub-mm observations should eventually be able to distinguish between 
these two possibilities, since the dust mass in the latter case would be 
substantially larger than for slightly dusty red cluster ellipticals.

In addition to the above EROs, we 
also found 4 objects which originally appeared to be EROs from the
$I-K$ color, but were found to be bright at $V$-band. Three of these
objects appear in the highest-$z$ field, MRC\,0527-255. These objects
appear similar to the {\it FROG} objects (Faint Infrared-Excess Field Galaxies)
discovered by Moustakas et al.~(1999), 
and may represent objects with a nuclear starburst or AGN.

%%%%%% ERO stuff:
%	1) how significant is ERO at edge?
%	2) could be more mergers at edge of field.
%
%
 
%The eros may also simply be function of cluster density, (ie different objects
%but tracing the fractional number of the underlying gal. density.)

%\subsection{Blue galaxy fraction}

%Motivation: higher fraction of blue galaxies at higher z than lower (B.O)

%And does the blue fraction hint at the evolutionary environment of the
%RG itself?

%Q: is the cluster redshift ambiguous? What can we say about color evolution

%we can look at intercept, slope well enough,

%we can look at scatter with the caveat that our sample is not
%morphologically selected, so excess scatter can always be interpreted
%as interlopers.

\section{Discussion}

\subsection{Comparison with other clusters}

The magnitude, spatial, and color distributions of the excess galaxy population
in these RG fields are all consistent with the excess being produced by
overdensities of galaxies at the RG redshifts.
Roughly speaking, the amplitude of the near-field excess corresponds to Abell
richness class $\sim$1 clusters,
and the far-field excess to Abell richness $\sim$1.5.
This is consistent with RGs often being located in large-scale galaxy overdensities
and occasionally in smaller-scale groups %``condensations'' 
within them.  However, the richness measurements should be
interpreted cautiously, %taken with a grain of salt,
as one galaxy at $z$=0 may typically 
correspond to several galaxies at $z$$\sim$1
due to mergers in the interim (Steinmetz \& White 1997, Stanford et al.~1997).

The objects within 60 arcsec of the radio galaxies at $K < 20$ do not
have colors which are significantly different from the no-evolution
Coma colors transformed to the redshift of the RG.
Because the uncertainty in the zeropoint of the Coma no--evolution
prediction is similar to the amount of expected color evolution in $J-K$
and $H-K$, we would not 
find strong evidence for passive evolution of the cluster galaxies even
without our uncertainties in fitting the red sequence cluster redshift.

The lack of redshift identifications for cluster members, and the large
potential for intervening structure at these relatively high redshifts, makes
it difficult to estimate color evolution. 
We have shown that at least for
MRC\,1022-299, the red sequence estimated redshift is very
close to the RG redshift itself. For the case of the more dispersed cluster,
MRC\,1139-285, and the nearest cluster, MRC\,0959-263, 
the red sequence estimated redshift is 
somewhat different from that of the RG ($\delta$z=0.08). 
Although the radio galaxy redshift is not necessarily the
best redshift for the cluster
(the RG is not clearly the BCM galaxy in any cluster),
a velocity dispersion corresponding to d$z$=0.08 would be relatively large
for clusters at our RG redshifts 
(Deltorn et al.~1997, van~Dokkum et al.~1999, Rosati et al.~1999).
It would therefore be surprising if the cluster redshift were systematically
offset from that of the radio galaxy.
In addition, the photometry and Coma zeropoint errors added in quadrature
(3$\sigma \sim$ 0.2 mag)
are similar to the measured color offsets.
Therefore we are reluctant to associate such minor color offsets from
the Coma no-evolution colors with excess or diminished cluster evolution.
MRC\,0527-255 has only
the red sequence estimated redshift to work from, and until
spectroscopic redshifts are obtained, nothing further can be gathered
about the color evolution.

%color evolution may be significant at least in lower z clusters.???????

The colors of most of the objects lying near the red sequence
are broadly consistent with the predictions of Bruzual \& Charlot
(1997) 
elliptical galaxy models for $z = 0.76, 0.78, 0.95$ and $1.28$ respectively,
with the following properties:
passive evolution of a 1 Gyr burst stellar population with solar metallicity
formed at $z = 5$ for $h=0.65$, q$_0 = 0.1$.  For our most distant
possible cluster (MRC~0527; $z=1.28$), such a model predicts $V-K =
7.0$, $I-K = 4.3$ and $J-K = 1.9$ for a galaxy age of 3.25 Gyr.  
It is
noteworthy that the same ``standard'' passive--evolution model was also found
to provide a reasonable fit to the average optical--IR colors of early--type
galaxies in the large sample of rich optical and X-ray selected
clusters at $0.3 < z < 0.9$ (Stanford et al.~1998).  
Ages less than $\sim$3 Gyr for the stellar populations
could be accommodated if the mean metallicity were greater than solar.  Also,
we have made no attempt to correct the colors for extinction due to dust
internal to the galaxies.  If this is large, then the true colors of the
cluster galaxies would be bluer than our measured values, resulting in a
better fit to models with $z_f < 3$.

The apparent scatter and slope of the red sequence can be used as a measure
of the relative mix of stellar populations present in these early
epoch cluster ellipticals.
Our lack of morphology selection from high spatial resolution (ie: HST)
images makes color scatter comparisons difficult, since we have inadvertently
removed all bluer galaxies from consideration as cluster members.
Stanford et al.~(1998) noted no significant increase in red sequence
scatter to $z\sim0.9$ using HST-selected E/S0 galaxies in clusters.
Beyond $z=1$, the mix of stellar populations is expected to increase
the scatter and slope 
(Kodama et al.~1998), leaving a clear excess of red galaxies
but a much looser ``sequence''. This appears to be the case in two
recent analyses of AGN-selected clusters: 3c\,324 at $z=1.206$ 
(Kajisawa et al.~2000) and B2\,1335+28 at $z=1.11$ (Tanaka et al.~1999).
However, the $z\sim1.3$ clusters discovered in the Lynx IR field survey,
RX\,J0848.9+4452 and CIG\,J0848+4453 (Rosati et al.~1999), still show
a relatively tight red sequence. 
%similar to that found by Stanford et al.~(1997).
Our apparent red sequence scatter and slope out to $z\sim0.9$ appear similar to the eye as that found by Stanford et al.~(1997), but we do not attempt to
quantify the slope or scatter as described in section 3.1. 
MRC~0527-255, possibly at $z\sim1.3$, does show a rather loose red
sequence similar to 3c\,324, 
although the pile up of red objects at fainter magnitudes is 
fraught with photometric errors, and it is not clear which objects to include 
in the sequence.

%[EVIDENCE THAT THE MIX OF STELLAR POPULATIONS APPROACHING Z=2 IS
%INHOMO ENOUGH THAT ALL YOU GET IS AN EXCESS OF RED GAL, BUT NO
%TIGHT SEQUENCE. CLEARLY THIS IS THE $z$ REGIME TO EXPLORE!]

%%QUES: how significant could it potentially be within phot errors?

\subsection{Connections with the radio galaxy}
%Evolution of the AGN in cluster environment

These four clusters are clearly amongst the highest density environments
in our total sample of high-$z$ radio galaxies (Chapman \& McCarthy,
in preparation). The radio power and source structure 
varies significantly through the four
RGs, even after accounting for redshift differences (Table~1).
However the spectral index ($\alpha_{843/408}$) is
similar for the four objects, placing them all slightly shallow of the
steep spectrum source definition ($\alpha>0.9$) from Kapahi et al.~(1998).
The relatively steep spectrum is typical for distant RGs (McCarthy 1993).
The RG is not clearly the BCM galaxy in any of the clusters (although
it is one of the two or three obviously brightest sequence members in
MRC~0527 and MRC~1139). This may be expected
if the formation and evolution of the radio galaxy depends for instance only
on a few trigger merger events early in the cluster formation period.
The radio galaxy properties would then only loosely
relate to the local environment.

Although early suggestions that nuclear properties and radio source
structure are linked to the clustering environment (Heckman et al.~1986), 
we find no obvious evidence for environmental differences in our 
cluster fields over those selected through X-ray or optical techniques.
Higher merger rates might be expected to associate with high$-z$ RG cluster
environments, especially those less rich clusters with smaller velocity
dispersions (e.g. Barnes 1999). 
However, direct studies of merger fractions are difficult without 
higher resolution imagery,
and blue fractions of clusters at high$-z$ are difficult to quantify
due to the strong variation in the blue field population in the foreground
(Ellis 1997).
% the lack of evolution in the measured scatter and slope of the red sequence
%colors alone, the effect of increased merger activity is not observed.
Our clusters do all show a compact environment present in the central
$\sim$35\arcsec\ which is somewhat distinct from richer X-ray selected clusters
at similar redshifts (van~Dokkum et al.~1999, Rosati et al.~1999).
This may indicate a locally higher merger environment
in the cluster center than typical (by contrast, van Dokkum et al.~1998
found higher merger fractions in the outer radii of the clusters through
HST imaging of close pairs).
It remains to be seen whether there is any true correlation of cluster
and radio properties in a large flux limited sample.

\section{Conclusions}

The observations presented here provide strong evidence supporting the
cluster identification for the very red galaxy overdensities
in the environments of these four radio galaxies
environments.
The density of all objects within 25\arcsec radius of the radio galaxy
field centers is $\sim6\times$ higher
than the field average. Excluding the bluer objects, the overdensity
peaks at $\sim7.5$ times the field count subjected to similar $I-K$ color cuts.

The quiescent behavior of early type galaxies in clusters
to $z\sim1$ appears to represent the evolutionary history of
these RG-selected clusters fairly well.
The smooth and steady
evolution picture has been demonstrated with only a handful of relatively
rich clusters
selected through the X-ray and optical morphology thus far, and
our four clusters add significant evidence to this picture.

No clear connection exists in our data between the radio galaxy properties 
and the clustering environment, except for the local peak in the overdensity
surrounding the radio galaxy.

%\noindent{$\bullet$} %We find that X 

\section*{Acknowledgements}

We thank the  staff of the Las Campanas Observatory, and F. Peralta and H. Olivares, in
particular, for their assistance with the observations. We also thank D. Murphy
for technical assistance with IRCAM, and M. Beckett, D. Murphy, and R. McMahon for
technical assistance with CIRSI.
The referee helped us to improve the final version by providing a valuable 
critique of the submitted version of this paper.

%\clearpage

%\newpage
%
% FIGURE 1
%
\begin{figure*}
\begin{minipage}{170mm}
\begin{center}
\epsfig{file=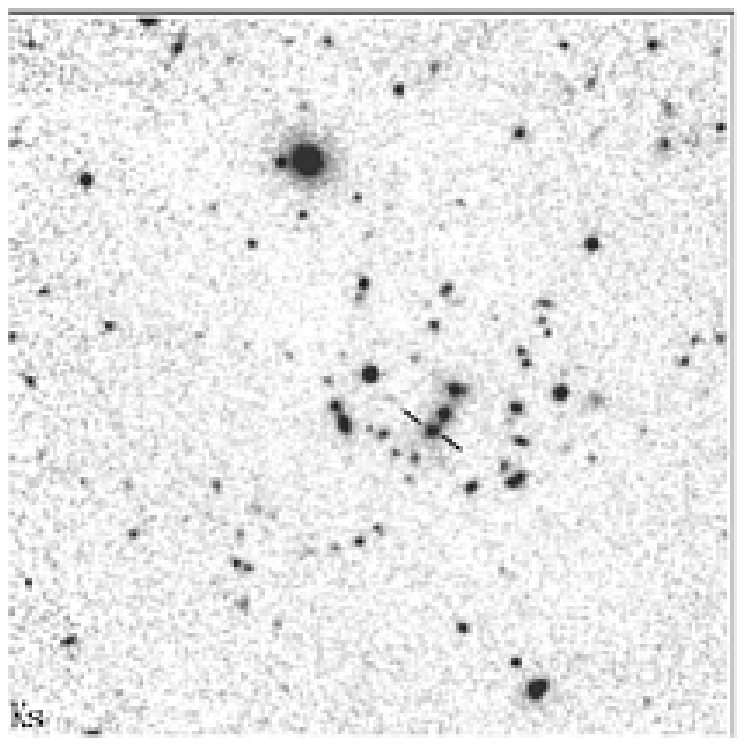,width=7.0cm,angle=0}
\epsfig{file=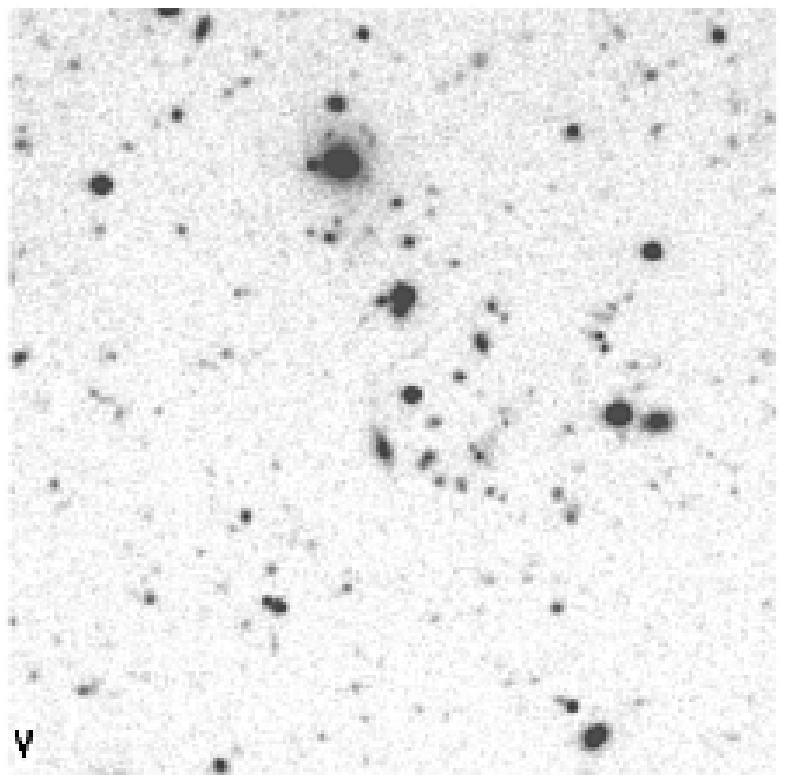,width=7.0cm,angle=0}
\epsfig{file=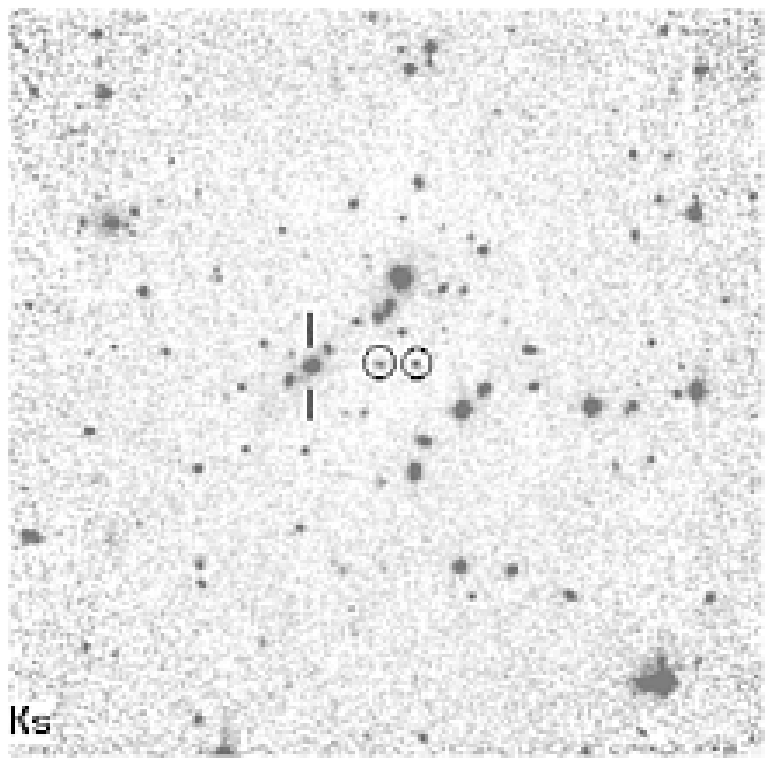,width=7.0cm,angle=0}
\epsfig{file=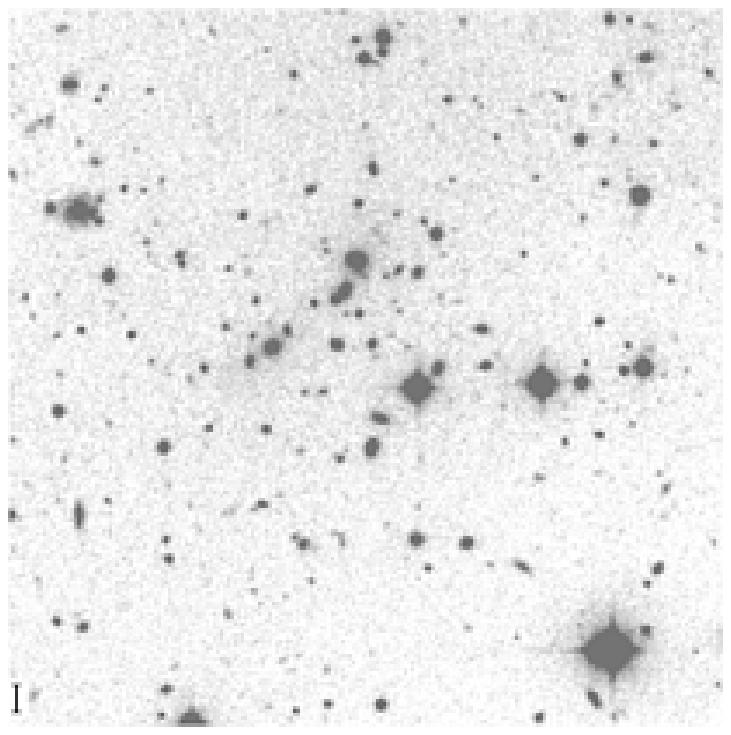,width=7.0cm,angle=0}
\end{center}
\figcaption[chapman.fig1]{Cluster field images comparing the near-IR to optical.
Top row: MRC~0527-255\,$K$, MRC~0527-255\,$V$; field size 95 arcsec on
a side.
Bottom row: MRC~0959-263\,$K$, MRC~0959-263\,$I$; field size 110 arcsec
on a side.
The radio galaxy is marked in the $K$s-band images.
The other cluster members with confirmed redshifts in MRC~0959-263
are circled.
In all fields in this paper, north is up, east is left.
\label{f1}}
\end{minipage}
\end{figure*}

%
% FIGURE 2
%
\begin{figure*}
\begin{minipage}{170mm}
\begin{center}
\epsfig{file=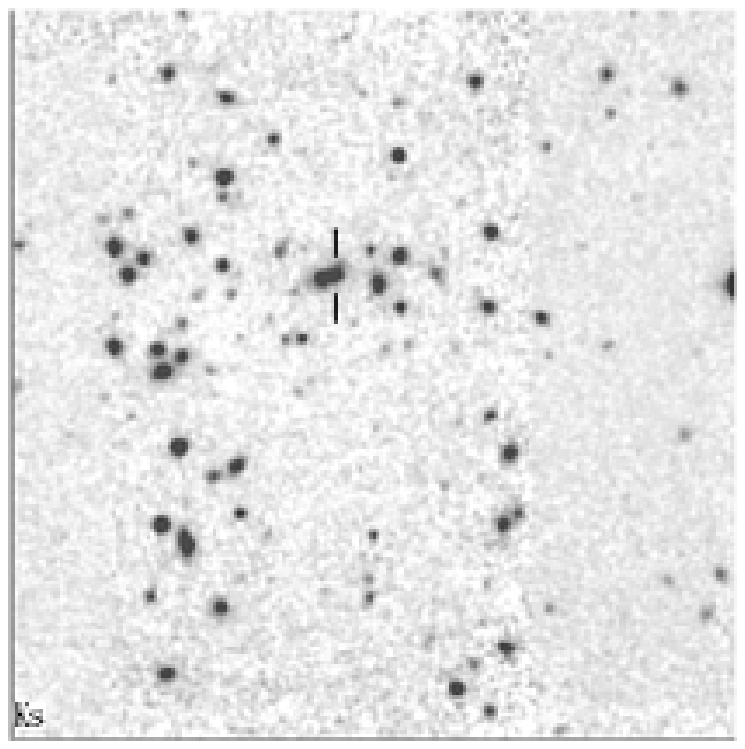,width=7.0cm,angle=0}
\epsfig{file=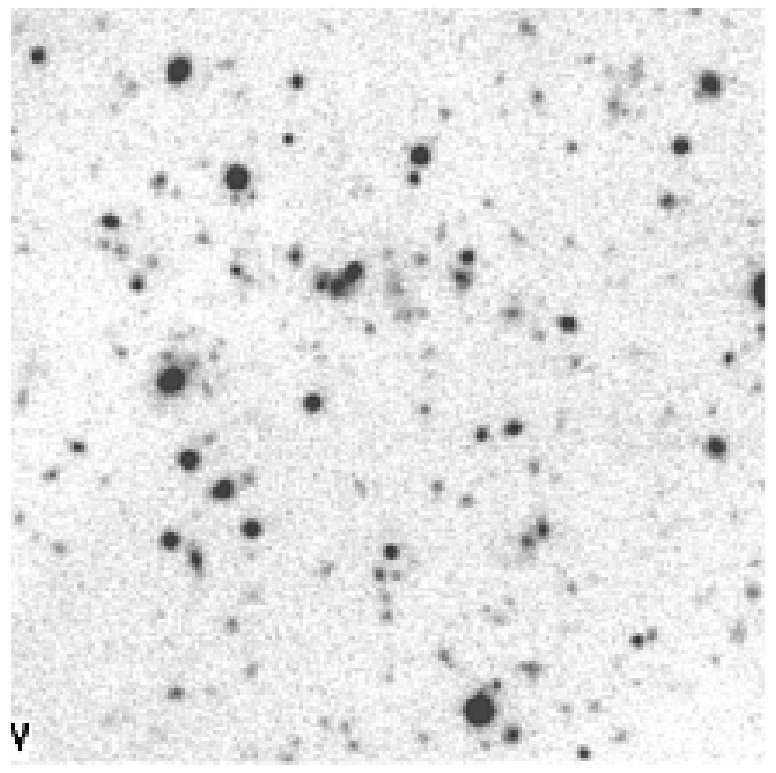,width=7.0cm,angle=0}
\epsfig{file=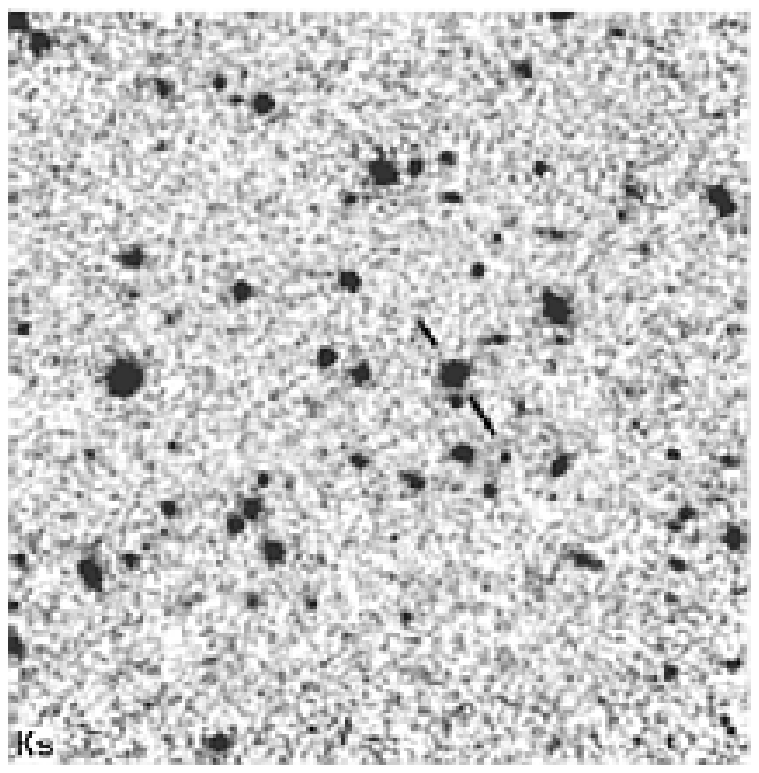,width=7.0cm,angle=0}
\epsfig{file=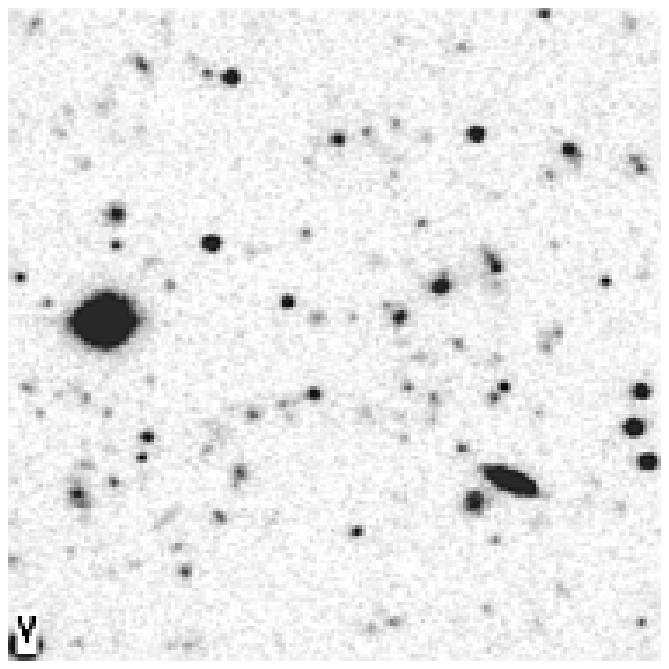,width=7.0cm,angle=0}
\end{center}
\figcaption[chapman.fig2]{Cluster field images comparing the near-IR to optical continued.
Top row: MRC~1022-299\,$K$, MRC~1022-299\,$V$; field size 90 arcsec on a side.
Bottom row: MRC~1139-285\,$K$, MRC~1139-285\,$V$; field size 90 arcsec on a
side.  The radio galaxy is marked in the $K$s-band images.
\label{f2}}
\end{minipage}
\end{figure*}

%
% FIGURE 3      MRC1022 false color
%
\begin{figure*}
\begin{minipage}{170mm}
\begin{center}
\epsfig{file=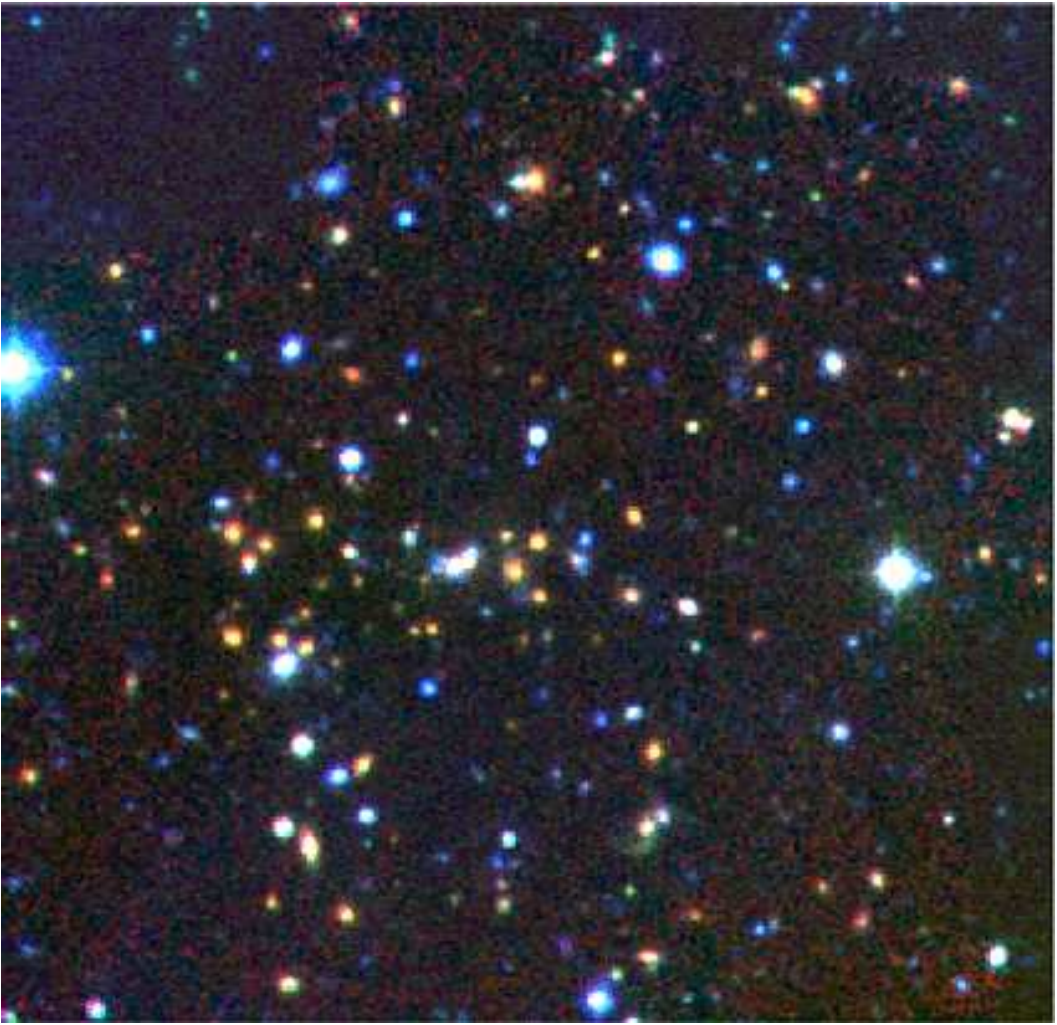,width=14.0cm,angle=0}
\end{center}
\figcaption[chapman.fig3]{$V$,$I$,$K$-band false color image of the richest
cluster in our sample (120 arcsec on a side),
MRC~1022-299. The excess red galaxy population clearly stands out in
an already overdense field.
\label{f3}}
\end{minipage}
\end{figure*}

%
% FIGURE -- position color (4)
%
\begin{figure*}
\begin{minipage}{170mm}
\label{poscol}
\begin{center}
\epsfig{file=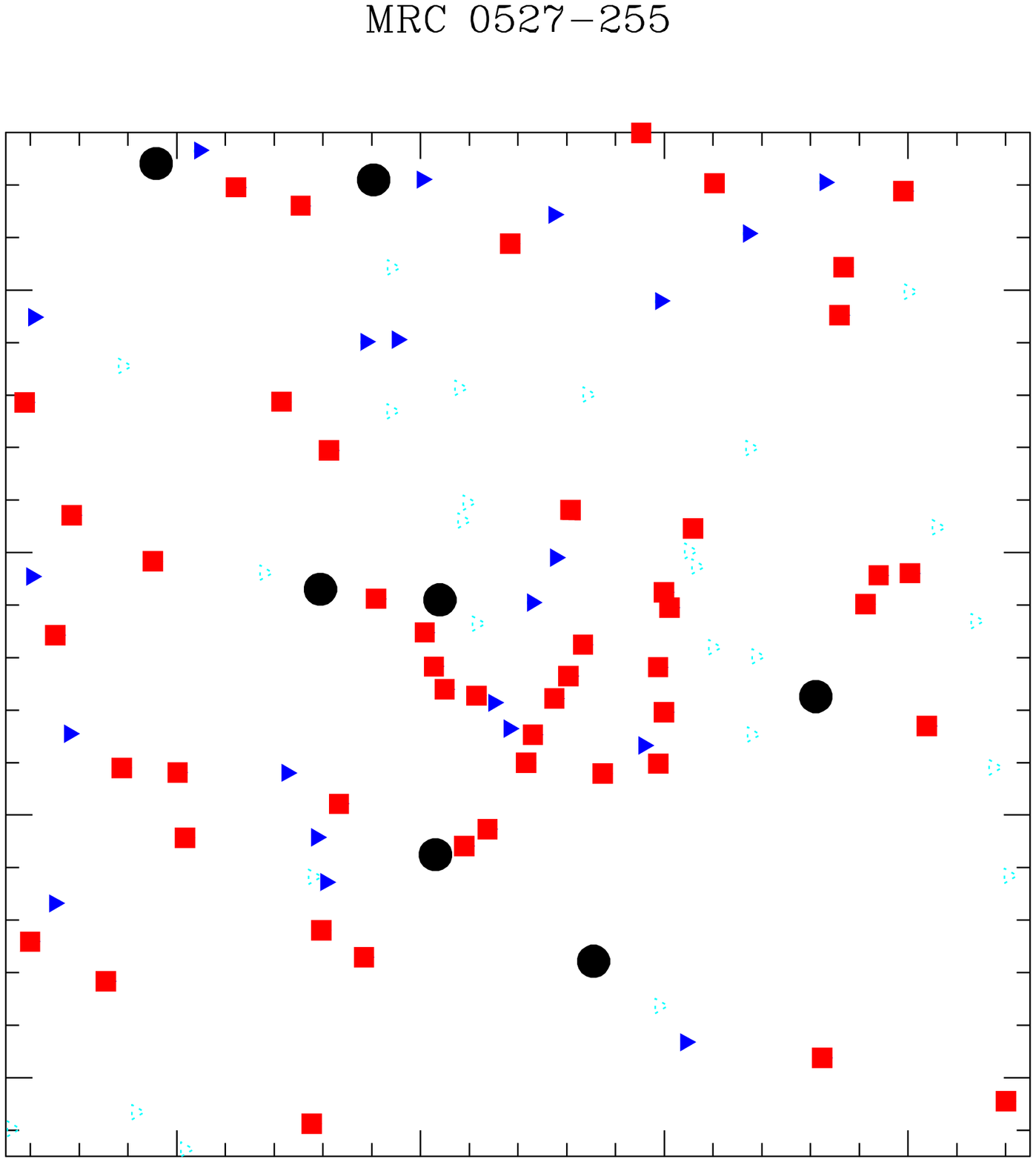,width=8.25cm,angle=-90}
\epsfig{file=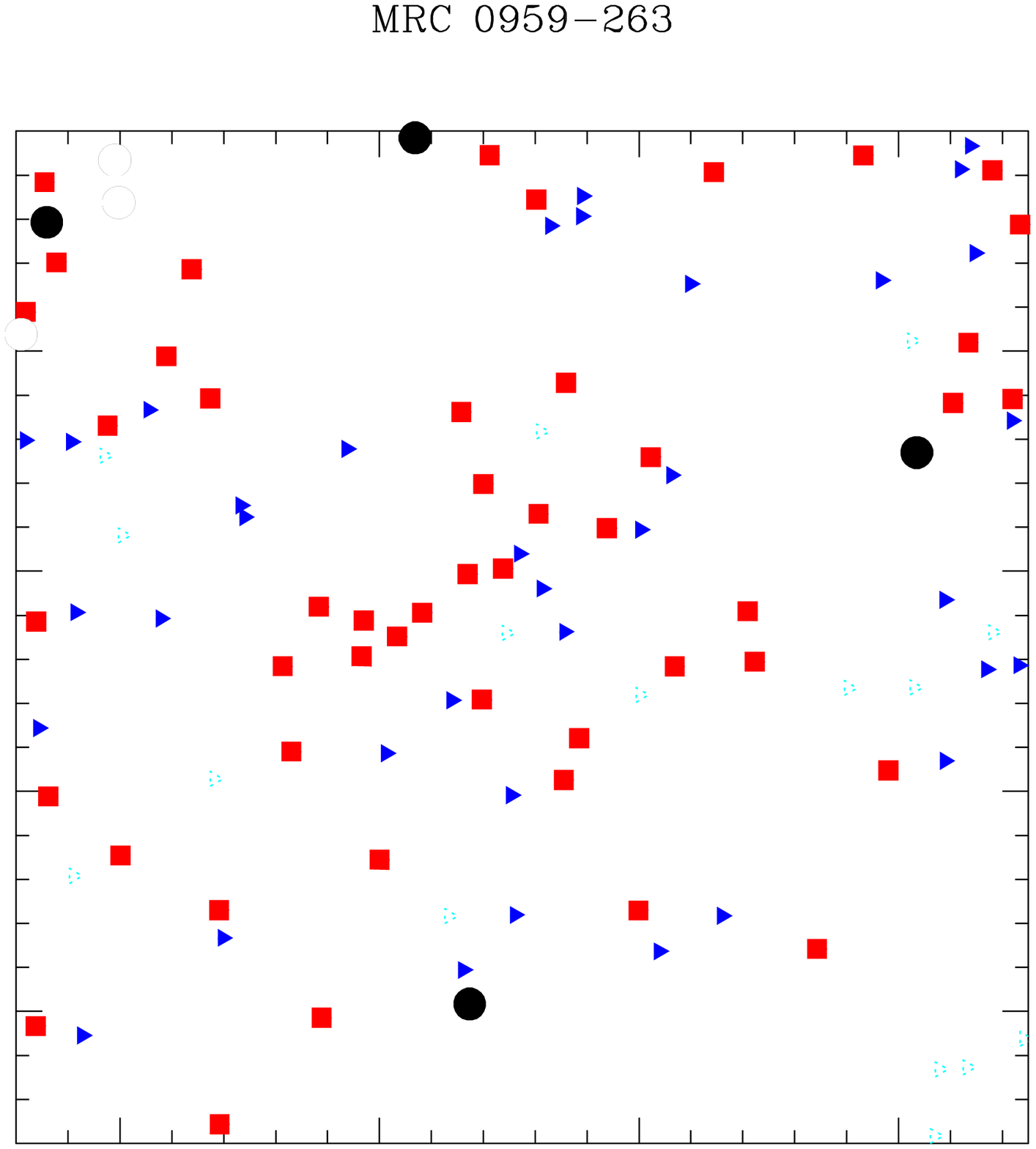,width=8.25cm,angle=-90}
\epsfig{file=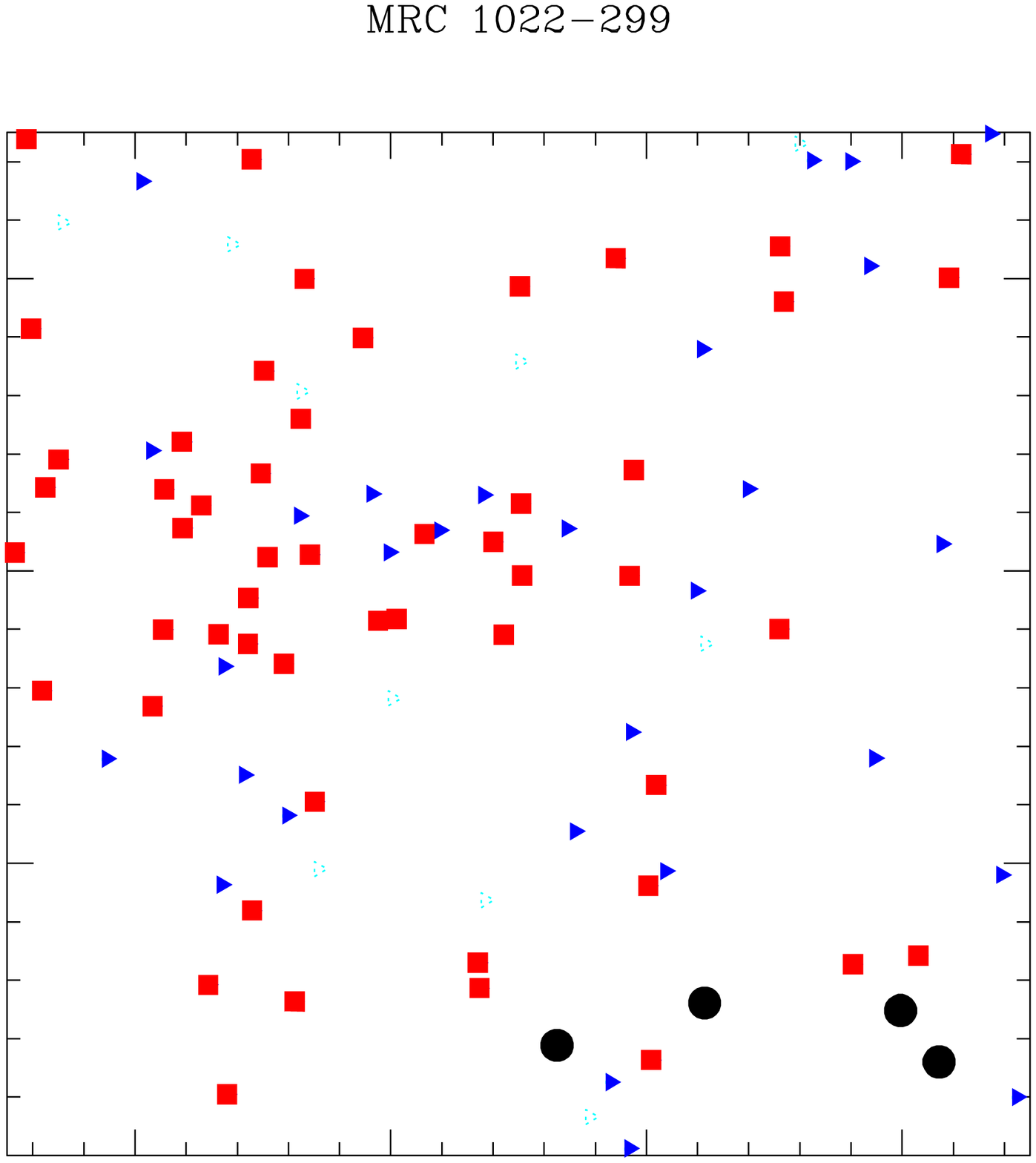,width=8.25cm,angle=-90}
\epsfig{file=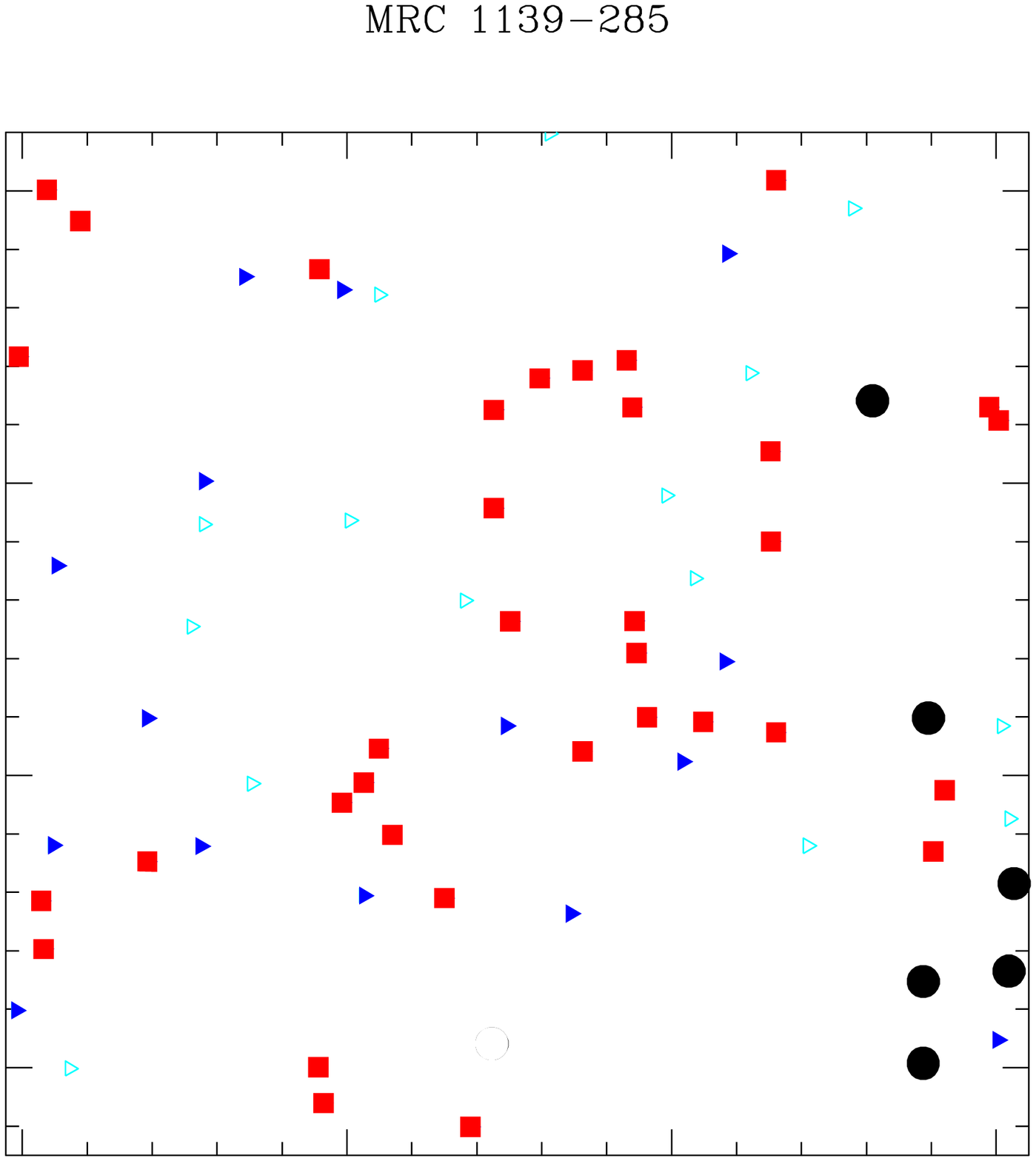,width=8.25cm,angle=-90}
\end{center}
\figcaption[chapman.fig4]{The 4 cluster position-color diagrams
showing the density of red sequence and ERO objects.
The angular extents of the plots are similar to the images in figures 1\&2.
Only objects with a $>4\sigma$ detection at $K$s-band are depicted.
Top row: MRC\,0527-255 and MRC\,0959-263
Bottom row: MRC\,1022-299 and MRC\,1139-285.
Solid red squares represent objects
lying on the red sequence, solid blue triangles represent
galaxies with $I-K>1.2$ mag.~bluer than the red sequence, and open blue
triangles
with $I-K>2$ mag.~bluer than the red sequence.
EROs with $I-K$ at least 1 mag.~redder than the respective
elliptical sequence are
shown with black circles (at least $I-K>4.3$).
\label{f4}}
\end{minipage}
\end{figure*}

%
% FIGURE CMDs (5)
%
\begin{figure*}
\begin{minipage}{170mm}
\label{fcmd}
\begin{center}
\epsfig{file=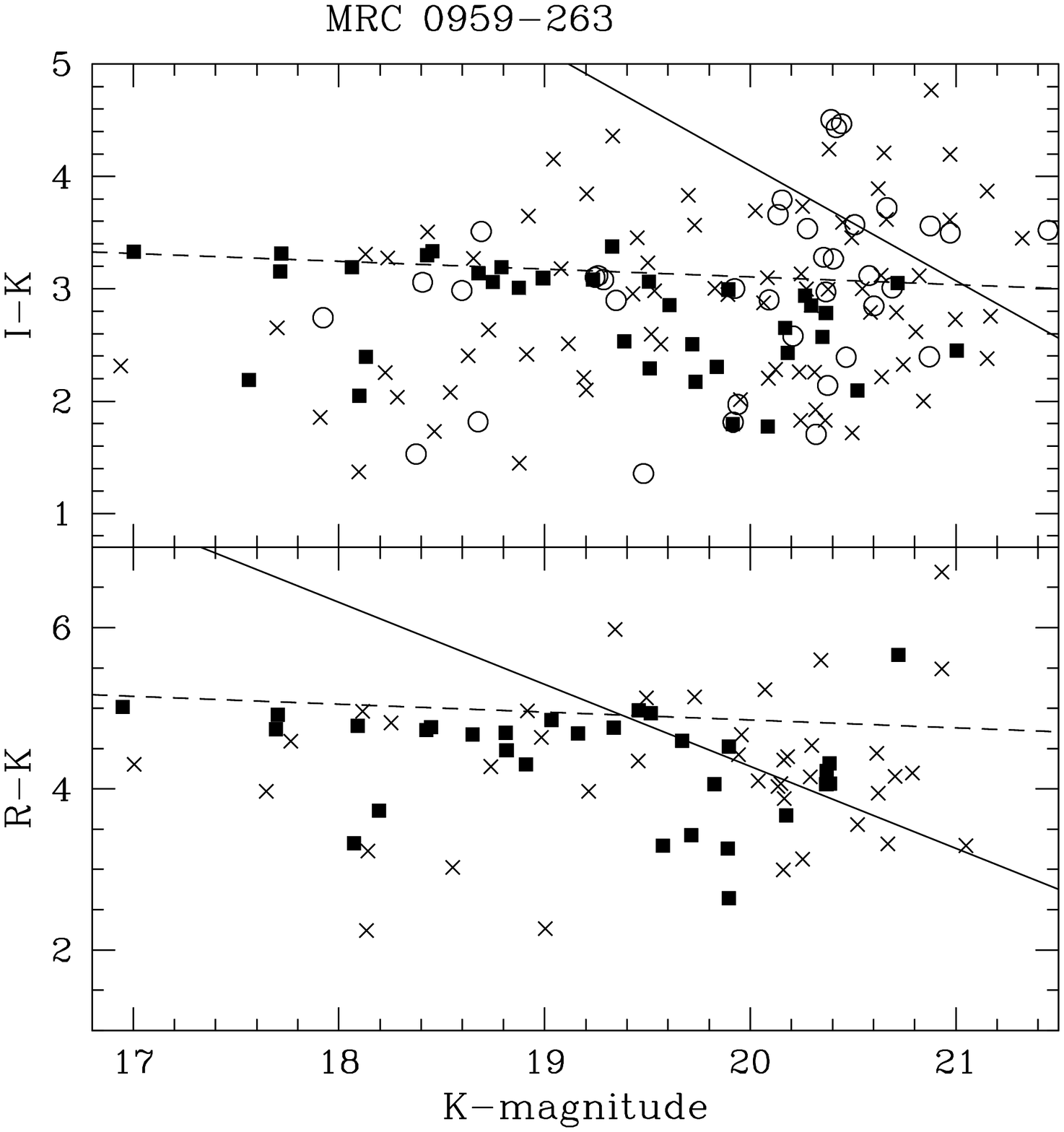,width=8.25cm,angle=0}
\epsfig{file=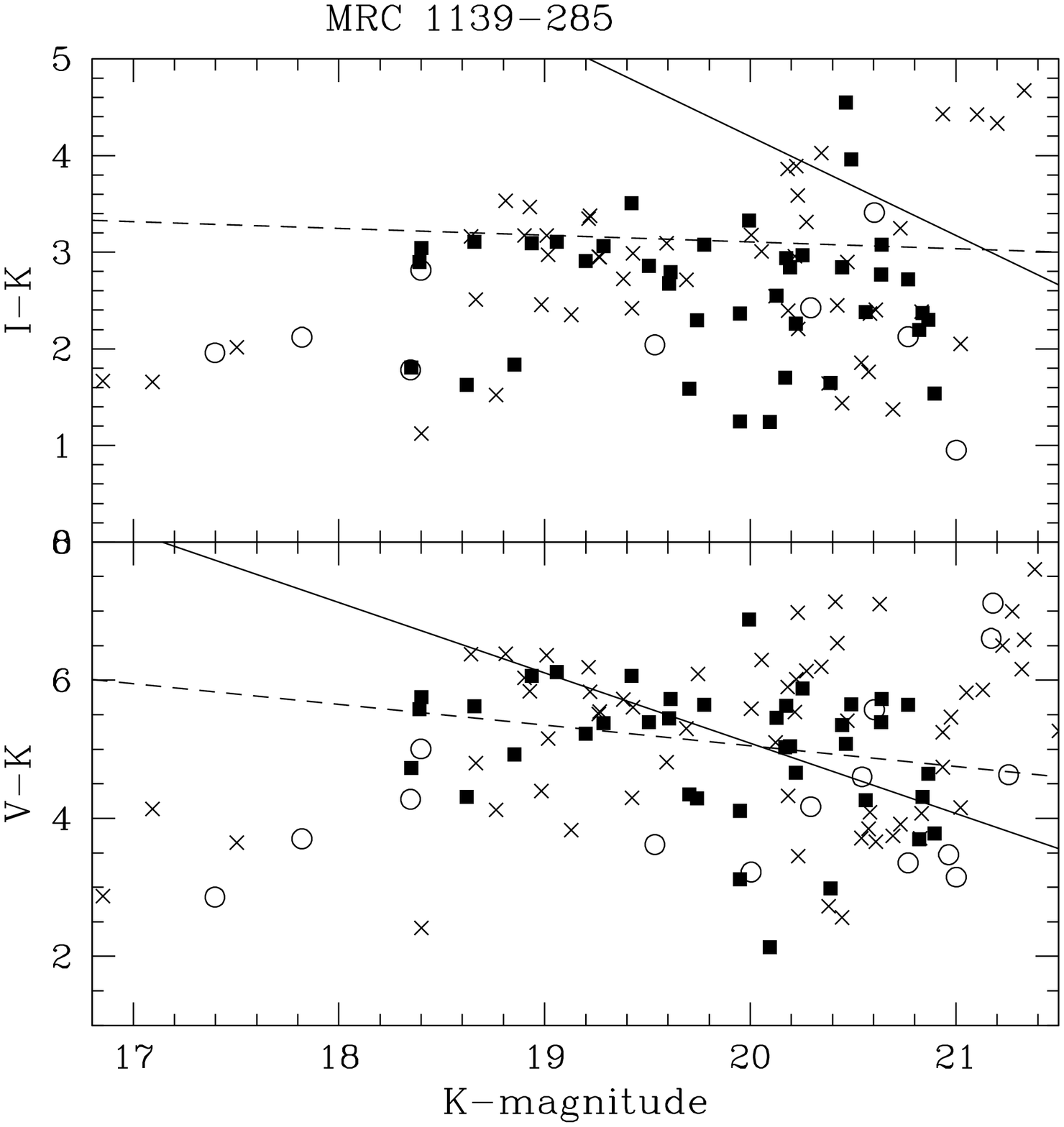,width=8.25cm,angle=0}
\epsfig{file=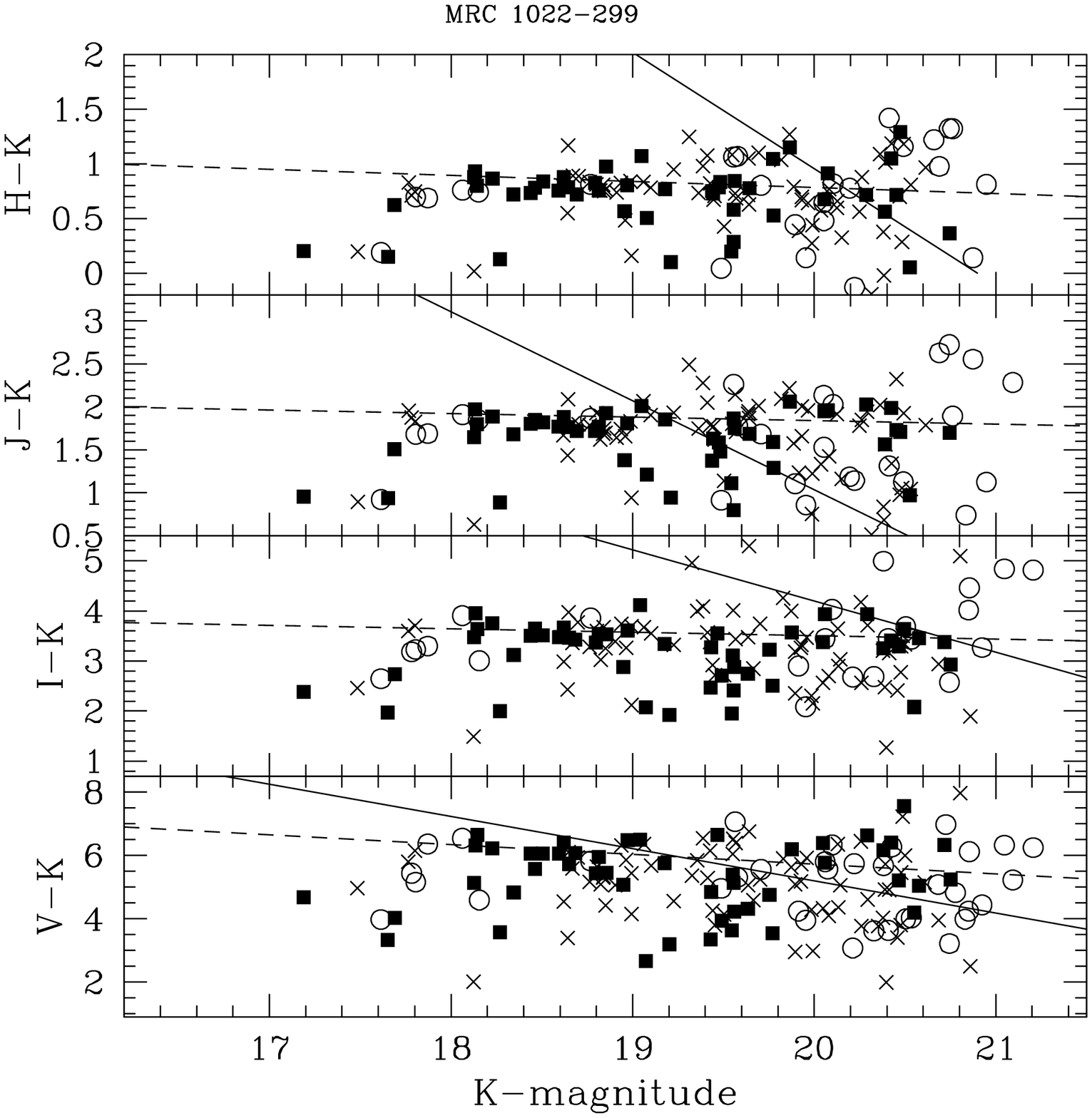,width=8.25cm,height=10cm,angle=0}
\epsfig{file=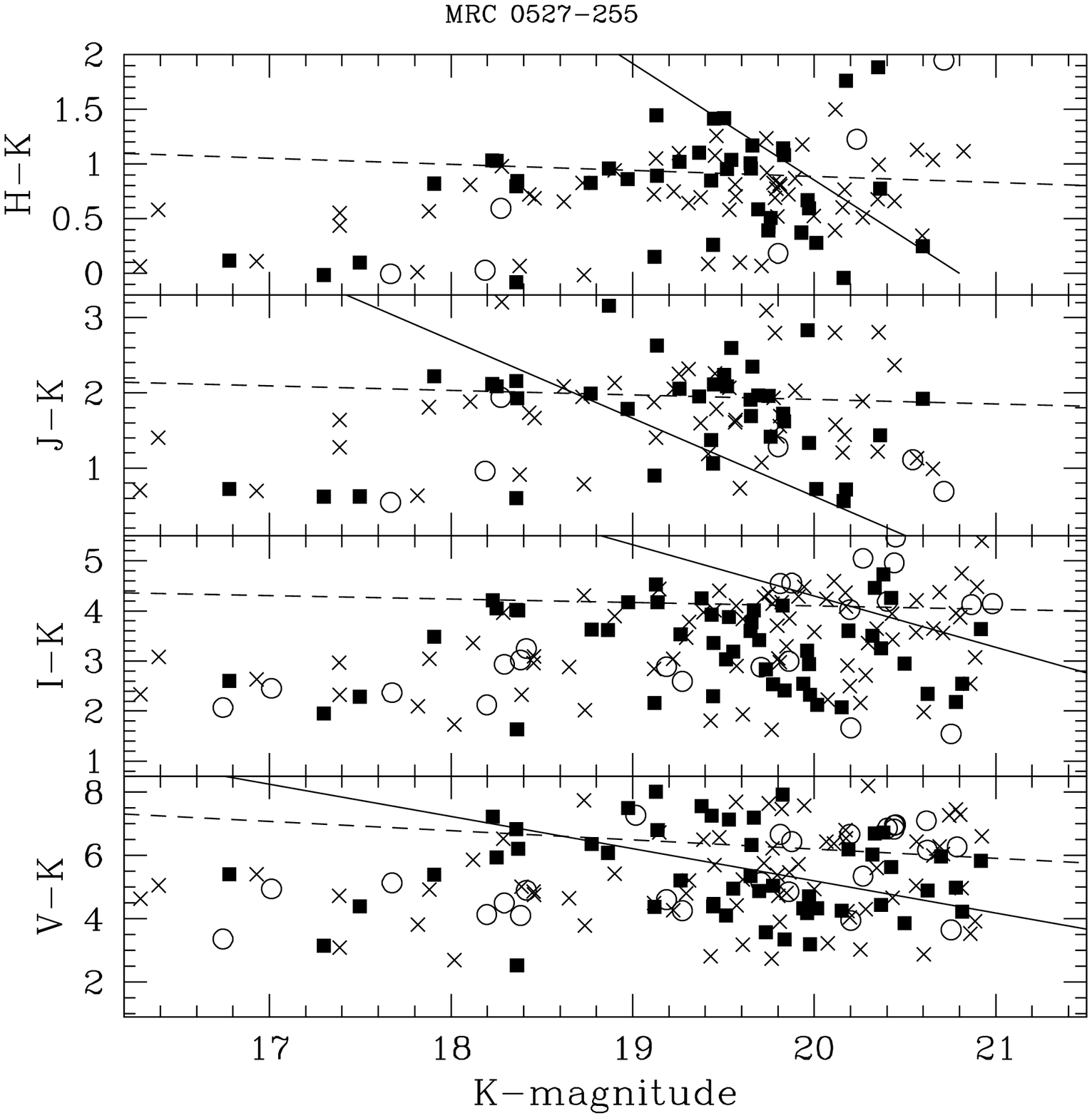,width=8.25cm,height=10cm,angle=0}
\end{center}
\figcaption[chapman.fig5]{The cluster CMDs at 2 radial cuts
showing the red sequence emerging.
In order of increasing redshift; Top row: MRC\,0959-263; MRC\,1139-285,
Bottom row: MRC\,1022-299; MRC\,0527-255. Radial cuts have been made on the 
objects to provide an idea of the color distributions with distance from the
radio galaxy. Solid squares represent objects
within 35 arcsec of the radio galaxy, crosses from 35\arcsec\ to 60\arcsec\,
and circles from 60\arcsec\ to 90\arcsec\,. Dashed lines depict the
locus of Coma cluster colors transformed to the redshift of the
RG-selected cluster. The solid diagonal lines represent $4\sigma$ detection
thresholds in color, with magnitude errors added in quadrature.
Note that for MRC\,0959-263, the
Gunn $r$-band image covers a somewhat smaller
area than the $I$ and $K$ images, and many of the galaxies identified in the
$I$,$K$ fields are not present in $r$.
\label{f5}}
\end{minipage}
\end{figure*}

%
% FIGURE 6 -- MRC0527 SED
%
\begin{figure*}
\begin{minipage}{170mm}
\label{0527photz}
\begin{center}
\epsfig{file=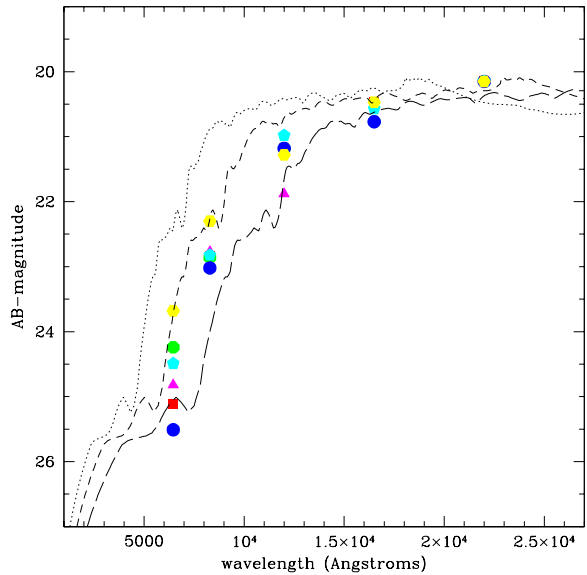,width=8.25cm,angle=0}
\end{center}
\figcaption[chapman.fig6]{MRC~0527-255:
A spectral energy plot of the six $K$-band brightest galaxies
lying along the
assumed red sequence for the cluster. The lines represent an elliptical
galaxy redshifted and k-corrected to $z=0.8$ (dotted), $z=1.0$ (dashed),
and $z=1.28$ (long dashed).
Although evolutionary effects will make this model SED less red, dust
effects would considerably redden the profile, providing a better fit to the
data at wavelengths shortward of 1~micron.
These six red objects provide some indication that the cluster likely
does lie at $z>1$.
\label{f6}}
\end{minipage}
\end{figure*}

%
% Figure 7
%       radial density
%
\begin{figure*}
\begin{minipage}{170mm}
\label{density}
\begin{center}
\psfig{file=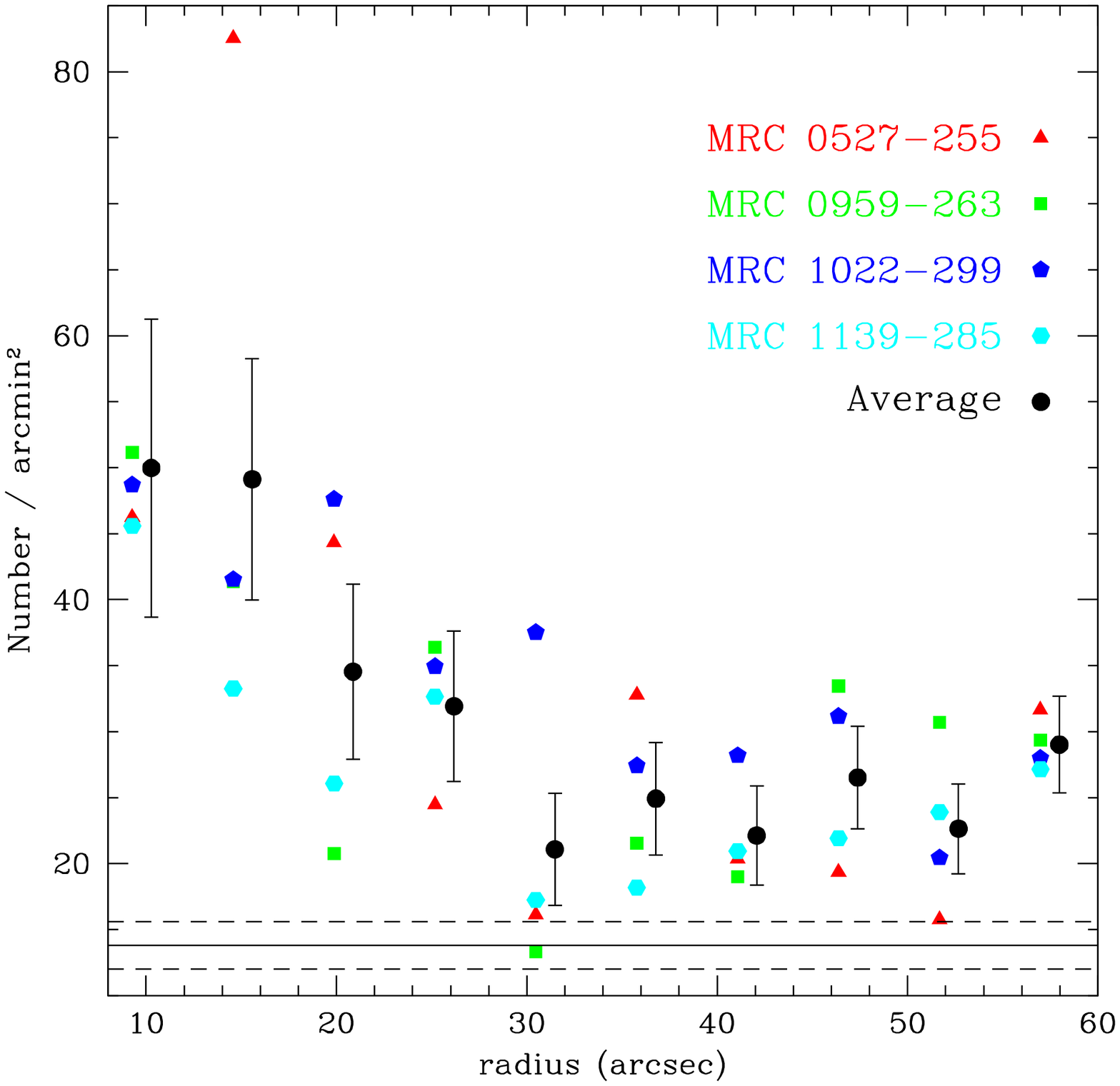,width=8.25cm,angle=0}
\psfig{file=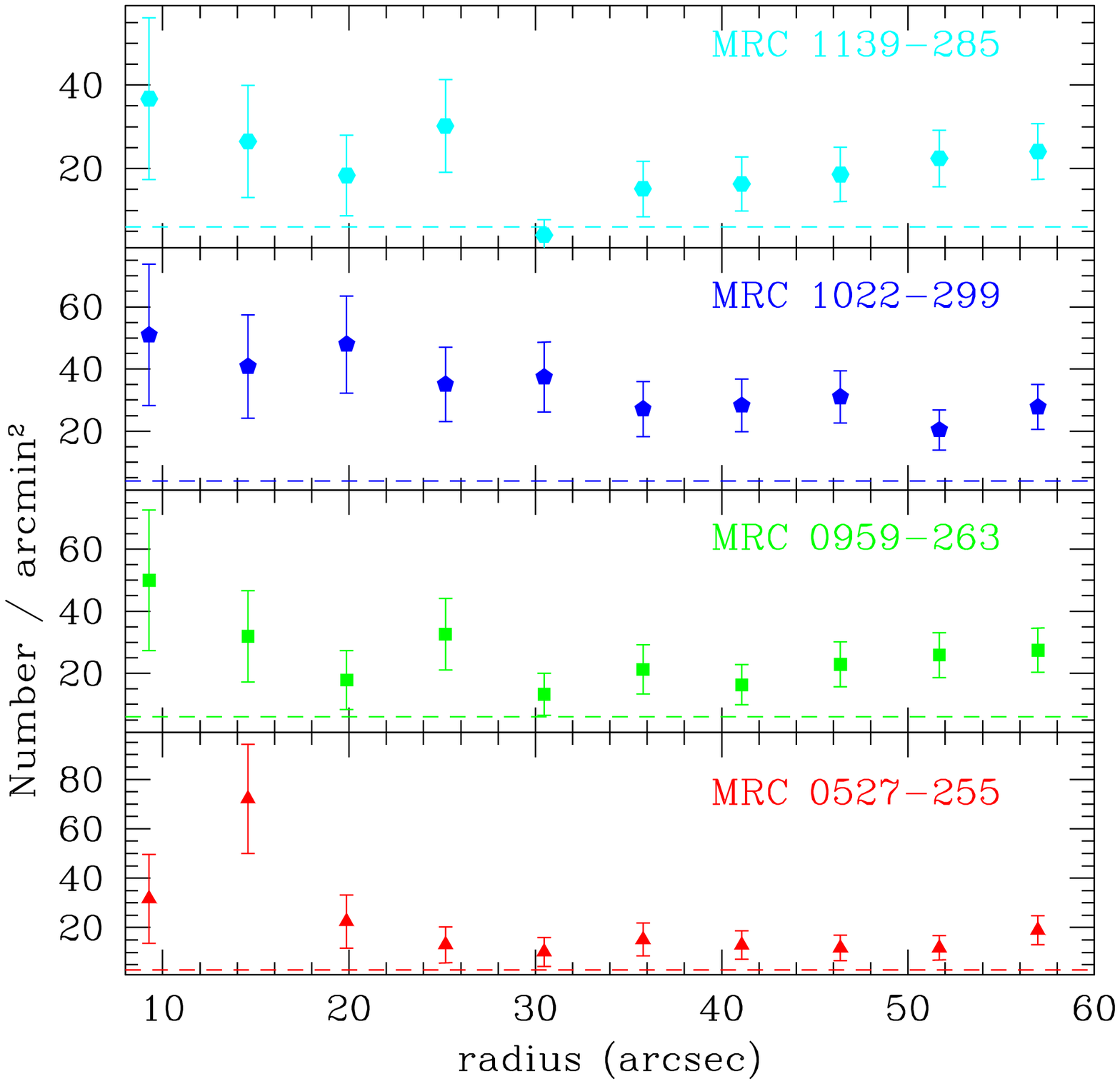,width=8.25cm,angle=0}
\end{center}
\figcaption[chapman.fig7]{Fig.~7a (left panel).
Projected radial distribution of all galaxies brighter than $K_s = 20.5$
(2\arcsec\ aperture diameter) within the
average of the 4 clusters (filled circles), with error bars are
calculated for the number of objects in each bin as per
Gehrels (1986).
The individual clusters are
also plotted, slightly offset in radius for clarity,
with MRC~0527 (triangles), MRC~0959 (squares),
MRC~1022 (pentagons) and MRC~1139 (hexagons).
A clear signal is seen as a significant excess above the field galaxy
limit at our $K$s-band survey depth (solid horizontal line, 1$\sigma$ error
bars as dashed lines).
%The figure shows separately the distribution of galaxies within
%a color cut and outside this color range, both normalized to their
%respective backgrounds.
Fig.~7b (right panel). By making a color cut on the objects of
1 magnitude bluer than the $I-K$ red
sequence in each case, the strength of the overdensity
signal becomes even larger, as we have cut out a portion of the
contaminating field galaxy population. The $I-K$ color-cut limits for the
field in each case are taken from Cowie et al.~1996 (horizontal dashed lines),
and vary from $\sim6$/arcmin$^2$ (MRC\,0959-263) to $\sim3$/arcmin$^2$
(MRC\,0527-255).
Again, the error bars are calculated for the number of bins
as per Gehrels (1986).
\label{f7}}
\end{minipage}
\end{figure*}

\newpage
%
% TABLE 1
%
\begin{deluxetable}{lcccc}
\tablewidth{480pt}
\scriptsize
\tablenum{1}
\label{table-1}
\tablecaption{\sc \small Properties of
the Radio-Galaxy Clusters\label{tab1}}
\tablehead{
\colhead{Property} & \colhead{MRC~0527-255} & \colhead{MRC~0959-263} & \colhead{MRC~1022-299} & \colhead{MRC~1139-285}  }
\startdata
%\multispan4{$4\sigma$ Detections  \hfil}\\
$K$s $4\sigma$ limit& 20.5 & 20.3 & 20.5 & 20.4 \\
$H$ $4\sigma$ limit& 21.2 & n/a & 21.3 & n/a \\
$J$ $4\sigma$ limit& 21.0 & n/a & 21.4 & n/a\\
$I$ $4\sigma$ limit& 24.6 & 24.4 & 24.5 & 24.5 \\
$V$ $4\sigma$ limit& 25.5 & 24.6$^a$ & 25.5 & 25.2 \\
RG redshift   & n/a  & $0.68\pm0.005$  & $0.93\pm0.005$ & $0.85\pm0.005$ \\
%\multispan7{$4\sigma$ Detections  \hfil}\\
red sequence redshift & $1.28\pm0.03$  & $0.75\pm0.02$  & $0.94\pm0.02$ & $0.77\pm0.02$ \\
\noalign{\smallskip}
Abell class   & 1.5  & 1.5  & 2 & 1 \\
%\noalign{\medskip}
\noalign{\smallskip}
No.~red galaxies, R$<$10\arcsec\   & 6  & 5  & 3 & 2 \\
No.~red galaxies, R$<$35\arcsec\   & 22  & 13  & 16 & 9 \\
No.~red galaxies, R$<$60\arcsec\   & 32  & 31  & 38 & 20 \\
\noalign{\smallskip}
ERO$_{I-K>5}$ to $K$=20.0(20.7) &  3(6) & 3(5)  & 2(3) & 2(6)   \\
ERO$_{>\rm{red seq + 1~mag}}$ & 4(10)  & 3(7)  & 4(5)   & 3(8) \\
\noalign{\smallskip}
S$_{408\rm MHz} (mJy)$        & 1540 & 1720 & 1120 & 6810 \\
$\alpha_{843/408}$ & 0.88 & 0.82 & 0.84 & 0.85 \\
radio structural type & partly resolved & triple source & partly resolved & double source \\
\enddata
\vspace*{-0.5cm}
\tablerefs{$a$) This object has gunn-$r$ magnitude rather than $V$
%$a$) Ivison et al.\ (1998) --
%$g$) Edge et al.\ (1999).
}
\end{deluxetable}

\end{document}